\documentclass[preprint,showpacs,preprintnumbers,amsmath,amssymb]{revtex4}
\usepackage{epsfig}
\usepackage{graphicx}
\usepackage{dcolumn}
\usepackage{bm}

\newcommand{\ord}{{\cal O}}
\def\beq{\begin{equation}}
\def\eeq#1{\label{#1}\end{equation}}
\def\eeqn{\end{equation}}
\newcommand\iden{\leavevmode\hbox{\small1\normalsize\kern-.33em1}}

\let\jnfont=\rm
\def\NPB#1,{{\jnfont Nucl.\ Phys.\ B }{\bf #1},}
\def\PLB#1,{{\jnfont Phys.\ Lett.\ B }{\bf #1},}
\def\EPJC#1,{{\jnfont Eur.\ Phys.\ Jour.\ C }{\bf #1},}
\def\PRD#1,{{\jnfont Phys.\ Rev.\ D }{\bf #1},}
\def\PRL#1,{{\jnfont Phys.\ Rev.\ Lett.\ }{\bf #1},}
\def\MPLA#1,{{\jnfont Mod.\ Phys.\ Lett.\ A }{\bf #1},}
\def\JPG#1,{{\jnfont J.\ Phys.\ G }{\bf #1},}
\def\CTP#1,{{\jnfont Commun.\ Theor.\ Phys.\ }{\bf #1},}
\def\JHEP#1,{{\jnfont JHEP \ }{\bf #1},}
\def\NPPS#1,{{\jnfont Nucl.\ Phys.\ Proc.\ Suppl.\ }{\bf #1},}

\begin{document}

\preprint{\parbox{1.2in}{\noindent ~~ }}

\title{\ \\[10mm] Higgs-pair Production in Littlest Higgs Model with T-parity}

\author{Lei Wang, Wenyu Wang, Jin Min Yang, Huanjun Zhang }

\affiliation{Institute of Theoretical Physics, Academia Sinica,
             Beijing 100080, China }

\begin{abstract}
The Higgs-pair production process at the CERN Large Hadron Collider (LHC),
which will provide a way to test the Higgs boson self-coupling, may be
sensitive to new physics.
In the framework of the littlest Higgs model with T-parity,
such Higgs-pair production can proceed through additional loop diagrams
and thus the production rate can be quite different from the
Standard Model (SM) prediction.
Our calculations show that, due to the loop contributions of both T-even
and T-odd quarks predicted in this model, the production rate can be
significantly enhanced relative to the SM prediction and also can be
larger than the production rate in the minimal supersymmetric model.
Also, we find that the T-odd quark contributions, which were ignored in
a previous study, are equally important compared with the T-even quark
contributions.
\end{abstract}


\maketitle

\section{Introduction}
To solve the fine-tuning problem of the Standard Model (SM), the
little Higgs theory \cite{ref1} was proposed as a kind of
electroweak symmetry breaking  mechanism accomplished by a naturally
light Higgs sector. The Higgs boson remains light, being protected
by the approximate global symmetry and free from one-loop quadratic
sensitivity to the cutoff scale. The littlest Higgs model
\cite{ref2} provides an economical approach which implements the
idea of the little Higgs theory. Most of the constraints from the
electroweak precision tests  on little Higgs models \cite{ref3} come
from the tree-level mixing of heavy and light mass eigenstates,
which would require raising the mass of the new particles to be much
higher than TeV scale and thus reintroduce the fine-tuning in the
Higgs potential \cite{ref4}. However, these tree-level contributions
can be avoided by introducing a discrete symmetry called T-parity
\cite{ref5}. In such a scenario, the top quark has a T-even partner
(denoted as $T$) and a T-odd partner (denoted as $T'$).  In addition,
some extra T-odd  fermions need to be also introduced
in order to make the model T-parity invariant.
These predicted new T-even and T-odd quarks will cause some
effects in various processes, especially the top quark and Higgs boson 
processes \cite{wang}, at collider experiments.
In this note we focus on the Higgs-pair production process at the
LHC, which may be sensitive to new physics.

The Higgs-pair production process  at the LHC will provide a way to
probe the Higgs boson self-coupling $\lambda$. With design
luminosity, it is possible for the LHC to establish that the SM
Higgs boson has a non-zero self-coupling and that
$\lambda$/$\lambda_{SM}$ can be restricted to a range of 0-3.7 at
$95\%$ confidence level if its mass is between 150 and 200 GeV
\cite{ref17}. Such Higgs-pair production process has been studied in
various new physics models \cite{ref6}. Recently, this process was
studied in the littlest Higgs model without T-parity  \cite{ref7}
and  with T-parity \cite{ref8}. However, the study in \cite{ref8}
only considered the contributions of T-even quarks but ignored the
effects of the T-odd fermions. As shown in some recent analyses
\cite{ref9,ref10,ref11}, the T-odd fermions can also cause some
interesting collider phenomenology and their effects cannot be
ignored. Given the popularity of the littlest Higgs model with
T-parity and also the importance of the Higgs-pair production at
the LHC as a probe of Higgs self-interaction, we in this note give a
complete calculation for the Higgs-pair production in the littlest
Higgs model with T-parity by considering the contributions of both
T-even and T-odd quarks.

This work is organized as follows. In Sec. II we recapitulate the
T-odd fermions and the top-quark sector of the littlest Higgs model with
T-parity. In Sec. III, we calculate the Higgs-pair production at the
LHC. Finally, we give our conclusion in Sec. IV.

\section{About the littlest Higgs model with T-parity}

\subsection{Fermion Sector}
The original Littlest Higgs model \cite{ref2} is based on a
non-linear sigma model describing the spontaneous breaking of a
global $SU(5)$ down to a global $SO(5)$ at an energy
scale $f\sim\ord(TeV)$. The vacuum expectation value (VEV) of an
$SU(5)$ symmetric tensor $\Sigma$ is proportional to \beq \Sigma_0
\,=\, \left(\begin{array}{ccc}
0& 0& \iden\\
0& 1& 0\\
\iden& 0& 0\\
\end{array}\right),
\eeq{sigma0}
where $\iden$ represents a unit $2\times 2$ matrix. The
low energy dynamics of non-linear sigma is described in terms of the
field
\beq \Sigma(x) \,=\, e^{i \Pi/f} \Sigma_0 e^{i \Pi^T/f} \,=\,
e^{2i \Pi/f} \Sigma_0
\eeq{sigma_def}
with
\beq
\Pi(x)=\sum_{a=1}^{14} \pi^a(x) X^a,
\eeq{pionmatrix}
where $\pi^a(x)$ are the Goldstone particles corresponding to 14 broken
generators $X^a$ for the $SU(5)\to SO(5)$ breaking.

To implement T-parity in the fermion sector, it requires the introduction
of the mirror fermions. For each SM lepton/quark doublet, under the
$SU(2)_1\times SU(2)_2$ gauge symmetry, two fermion doublets $q_1(2,1)$ and
$q_2 ({1,2})$ are introduced. They can be embedded into incomplete
representations $SU(5)$ multiplets $\Psi_1$ and $\Psi_2$.
A right-handed $SO(5)$ multiplets $\Psi_R$ transforming
nonlinearly under the full $SU(5)$ is introduced to give mass
to the extra fermions.  The field content can be expressed as
\begin{equation}
\begin{array}{ccc}
\Psi_1=\left(\begin{array}{c} q_1 \\ 0 \\ 0_2 \end{array}\right)\,,
& \Psi_2=\left(\begin{array}{c} 0_2 \\ 0 \\ q_2
\end{array}\right) \,,&
\Psi_R=\left(\begin{array}{c} \psi_R \\ \chi_R \\ \tilde{\psi}_R
\end{array}\right),
\end{array}
\end{equation}
with
\begin{equation}
q_{1}=\left( \begin{array}{c} i d_{L_1}\\  -i u_{L_1} \end{array}\right),~~
q_{2}=\left( \begin{array}{c} i d_{L_2}\\  -i u_{L_2}\end{array} \right),~~
\tilde{\psi}_R=\left(\begin{array}{c} i d'_{R}\\ -i u'_{R} \end{array}\right)
\end{equation}
The first component of $\psi_R$ is irrelevant to our study (as shown later)
and the second component of $\psi_R$ is $ -iq_R$. The mirror fermions
can be given $\ord(f)$ masses via a mass term
\cite{ref5,ref10,ref11,ref12,ref13}
\begin{equation}\label{heavyyuk}
{\cal L}_{\kappa}=-\kappa_{ij}  f (\bar{\Psi}^i_2 \xi
+\bar{\Psi}^i_1 \Sigma_0 \Omega \xi^\dagger \Omega)\Psi^j_R +h.c. ,
\end{equation}
where $\xi=e^{i\Pi/f}$, $\Omega \equiv {\rm diag}(1,1,-1,1,1)$ and
$i,j=1,2,3$ are the generation indices. For simplicity we assume
the flavor diagonal and universal $\kappa$ in our study.

They transform under the $SU(5)$ as
\beq
\Psi_1 \rightarrow V^* \Psi_1\,, \hspace{.2in} \Psi_2 \rightarrow V
\Psi_2\,, \hspace{.2in}\Psi_R \rightarrow U\Psi_R, \hspace{.2in}
\xi \rightarrow V\xi U^\dagger, \hspace{.2in}
\Sigma \rightarrow V\Sigma V^{\rm T} ,
\eeq{su5}
where $V$ is an $SU(5)$ rotation matrix, $U$ is the unbroken $SO(5)$
rotation and is a non-linear representation of the $SU(5)$.
Under T-parity the transformations are defined as
\beq
\Psi_1 \leftrightarrow -\Sigma_0 \Psi_2,  \hspace{.2in} \Psi_R
\rightarrow -\Psi_R,  \hspace{.2in} \xi \rightarrow \Omega
\xi^\dagger \Omega.
\eeq{su6}
Thus $q_1\leftrightarrow -q_2$  and
$\Sigma \rightarrow \Sigma_0 \Omega \Sigma^\dagger \Omega \Sigma_0$
under T-parity. Following the above transformation,  the Lagrangian
is T-invariant.

The Lagrangian in Eq.(\ref{heavyyuk}) contains new Higgs boson
interactions and the mass terms for the T-odd fermions
\begin{eqnarray}
{\cal L}_{\kappa}
&\simeq&-\sqrt{2} \kappa f
\left[
\bar{d}_{L_-} d'_{R}+\frac{1+c_\xi}{2} \bar{u}_{L_-}
u'_R
-\frac{s_\xi}{\sqrt{2}}\bar{u}_{L_-} \chi_R
-\frac{1-c_\xi}{2} \bar{u}_{L_-} q_R
 \right]+{\rm h.c.} ,
\label{Kappa_int}
\end{eqnarray}
where we ignored the generation indices, and
$c_\xi =\cos\frac{v+h}{\sqrt{2}f}$ and
$s_\xi =\sin\frac{v+h}{\sqrt{2}f}$ come from the
non-linear sigma model field $\xi$, with $h$ and $v$ being the neutral
Higgs boson field and its vev, respectively \cite{ref11}.
The mirror fermion $u_{L_{-}} = (u_{L_1}+ u_{L_2})/\sqrt{2}$
is T-odd, and $u_{L_{+}} = (u_{L_1}- u_{L_2})/\sqrt{2}$ is
T-even and massless. The same definitions also apply to the
down-type mirror quarks. The
fermions $q_R$ and $\chi_R$ can obtain large Dirac masses
by introducing additional fermions, as described in detail in
\cite{ref5,ref12}. We also assume the Dirac mass terms $-m_q
\bar{q}'_L q_R-m_\chi \bar{\chi}'_L \chi_R$. From Eq.(\ref{Kappa_int})
we can see that the first component of the doublet $\psi_R$
does not appear and the T-odd down-type quarks have no tree-level
couplings with the Higgs boson. After diagonalizing the mass matrix,
we get the mass eigenstates $u_-$ , $\chi$ and $q$, which couple with
$h$ and $hh$, respectively.

\subsection{Top-quark Yukawa couplings}

In order to cancel the quadratic divergence of the Higgs mass
induced by top quark, it requires completing $Q_1$ and $Q_2$
multiplets for the third generation to representations of the
$SU(3)_1$ and $SU(3)_2$  subgroups of the full $SU(5)$:
$Q_1=(q_1,U_{L_1},0_2)^{\rm T}$ and $Q_2=(0_2,U_{L_2},q_2)^{\rm T}$.
In addition to the SM right-handed top quark field $u_R$, one must
also introduce additional singlets  $U_{R_1} $ and $U_{R_2}$.

For the top-quark Yukawa couplings, one can write down the following
Lagrangian \cite{ref5,ref10,ref11,ref12,ref13}
\begin{eqnarray}
{\cal L}_t &=& -\frac{\lambda_1}{2\sqrt{2}}f\epsilon_{ijk} \epsilon_{xy}
\left[(\bar{Q}_1)_i \Sigma_{jx} \Sigma_{ky}-
(\bar{Q}_2 \Sigma_0)_i \tilde{\Sigma}_{jx} \tilde{\Sigma}_{ky}
\right] u_R \nonumber \\
&& -\lambda_2 f (\bar U_{L_1} U_{R_1}+\bar U_{L_2} U_{R_2}) +{\rm h.c.} ,
\label{top_yukawa_int}
\end{eqnarray}
where the indices $i,j,k$ run from 1 to 3 whereas $x,y=4,5$.
Note that under T-parity these fields transform as
\begin{eqnarray}
Q_1 \leftrightarrow -\Sigma_0 Q_2,
~~U_{R_1}\leftrightarrow -U_{R_2},
~~u_R\rightarrow u_R.
\end{eqnarray}
Therefore, the T-parity eigenstates are defined as $U_{L_{-}} =
(U_{L_1}+ U_{L_2})/\sqrt{2}$ (T-odd), $U_{L_{+}} = (U_{L_1}-
U_{L_2})/\sqrt{2}$ (T-even), and the same definitions also apply
to the right-handed singlets. From the above Lagrangian we can get the
following Higgs boson interactions and the mass terms for fermions
\begin{eqnarray}
{\cal L}_t &\simeq&
-\lambda_1 f \left(
\frac{s_\Sigma}{\sqrt{2}} \bar{u}_{L_+} u_R
+\frac{1+c_\Sigma}{2} \bar{U}_{L_+} u_R
\right)
-\lambda_2 f \left(\bar{U}_{L_+} U_{R_+}+\bar{U}_{L_-} U_{R_-}
\right)+{\rm h.c.}\, ,
\label{top-yukawa}
\end{eqnarray}
where $c_\Sigma=\cos\frac{\sqrt{2}(v+h)}{f}$ and
$s_\Sigma=\sin\frac{\sqrt{2}(v+h)}{f}$ come from
the non-linear sigma model field $\Sigma$ \cite{ref11}.
The T-odd Dirac fermion $T'$ ($T'_L \equiv U_{L_-},~T'_R \equiv U_{R_-}$)
obtains a mass $m_{T'}=\lambda_2 f$, and has no tree-level coupling
with the Higgs boson. The left-handed (right-handed) top quark and
T-even T-quark are linear combinations of $u_{L+}$ and $U_{L+}$
($u_{R+}$ and $U_{R+})$.
After diagonalizing the mass matrix in Eq. (\ref{top-yukawa}),
we can get the  mass eigenstates $t$ and $T$ as well as their couplings
with the Higgs boson.

\begin{figure}[tb]
\begin{center}
\epsfig{file=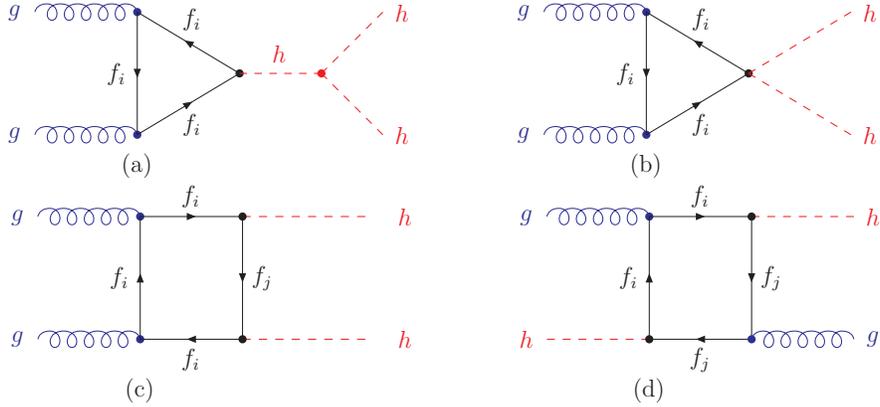,width=12cm}
\end{center}
\vspace{-1.0cm}
\caption{The parton-level Feynman diagrams for
 Higgs-pair production via gluon-gluon fusion in the littlest
Higgs model with T-parity. Here $f_i$ can be a T-even fermion
($i=1, 2$ with $f_1=t$ and $f_2=T$) or a T-odd fermion
($i=1, 2 ,3$ with $f_1=u_-$, $f_2=\chi$ and $f_3=q$).
The diagrams obtained by exchanging the two gluons or
exchanging the two Higgs bosons are not shown here.}
\end{figure}

\section{Higgs-pair production at LHC}

Now we look at the Higgs pair production  in the littlest
Higgs model with T-parity at the LHC. The production can proceed
through gluon-gluon fusion and $b\bar{b}$ annihilation at parton level,
with the former being the dominant one \cite{ref7}.
The Feynman diagrams of
Higgs-pair production via gluon-gluon fusion are shown in Fig. 1.
In the SM the dominant contributions are from the diagrams of
Fig.1(a, c, d) with top-quark loops.
In the littlest Higgs model with T-parity, the top-quark loops
give additional contributions through the tree-level $hht\bar t$
coupling and the modified $ht\bar t$ coupling. In addition to the
top-quark loops, the loops of new T-even and T-odd quarks
also come into play. So all these particles should be summed over
in our loop calculations. (As we pointed earlier, the calculations
in \cite{ref8} did not include the contributions of T-odd quarks).

The calculations of the loop diagrams in Fig. 1 are straightforward.
Each loop diagram is composed of some scalar loop functions \cite{Hooft}
which are calculated by using LoopTools \cite{Hahn}.
The calculations are tedious and the analytical expressions
are lengthy, which are not presented here.

We numerically checked our results by comparing our $gg \to hh$
parton cross section with Ref. \cite{ref7}.
The calculations in \cite{ref7} considered the loop effects of
(i) the top-quark and T-even T-quark, (ii) the heavy neutral triplet
Higgs boson $\Phi^0$, and (iii) the first and second generation quarks.
Since the dominant contributions are from (i) \cite{ref7},  
their results should be in approximate agreement with ours if we only keep 
the contributions of the top-quark and T-even T-quark with the same input 
parameters.
We made such a comparison in Table 1. We see that our results
agreement quite well with \cite{ref7}.

\begin{table}[htb]
\caption{\small  The comparison between our results with  \cite{ref7}
for the contributions of the top-quark and T-even T-quark to
$gg \to hh$ cross section by using the same parameters
and the same Feynman rules.}
\centering
\begin{tabular}{|c|c|c|c|c|c|c|c|c|c|c|c|c|}
\hline
$\sqrt{\hat{s}}(GeV)$ & 350&400&440 &500&520 &540&700 &900 &1000&2000\\ \hline
$\hat{\sigma}(gg \to hh)$ (ours)
 &0.0902& 0.3459&0.4514& 0.4953& 0.4922&
       0.4841  & 0.3616  &0.2385   & 0.1985 & 0.0636\\  \hline
$\hat{\sigma}(gg \to hh)$ (in \cite{ref7})
&  0.0947  &  0.3595 &  0.4685
 &  0.5138 &  0.5107 &  0.5022&  0.3754&  0.2478 &  0.2063
&  0.0659 \\
\hline
\end{tabular}
\end{table}

Note that in the littlest Higgs model with T-parity, 
T-parity forbids the generation of a vev for the
triplet scalar field and also forbids the contributions of the new T-odd
particles to processes with external SM fermions at tree-level.
Therefore, the electroweak precision constraints on the model with T-parity
are generically quite weak and, as a result, the symmetry breaking scale
$f$ may be as low as 500 GeV \cite{ref14}.
When expanding in the power series of $v/f$,
we need to keep some higher orders since $v/f$ may be not so small (
for example, $v/f\approx 0.5$ for f=500 GeV).
Therefore, when expanding the $c_\Sigma$ and $s_\Sigma$ to
diagonalize the mass matrix in Eq. (\ref{top-yukawa}),
we keep the order up to $\ord(v^5/f^5)$.
The diagonalization of the mass matrix in Eq. (\ref{top-yukawa})
was performed numerically in our analyses (in
\cite{ref10,ref12} the approximate expressions are given).

\begin{figure}[tb]
\begin{center}
\epsfig{file=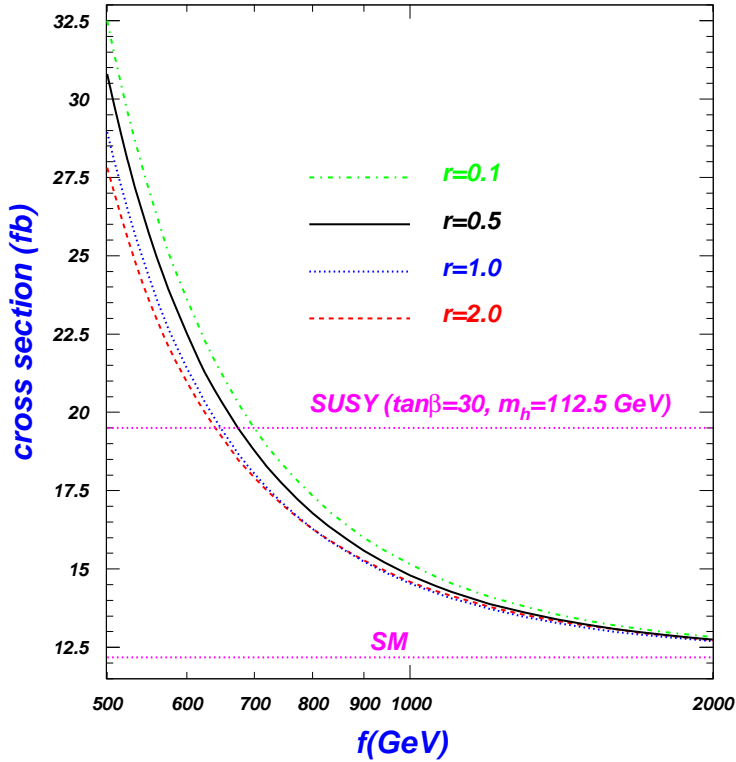,width=10cm}
\end{center}
\vspace{-1.0cm}
\caption{Hadronic Higgs-pair production cross section at the LHC
 versus the parameter $f$. The supersymmetric model prediction
is taken from Fig. 5(a) in the first reference of \cite{ref6}.}
 \end{figure}

The hadronic cross section at the LHC is obtained by convoluting the
parton cross section with the parton distribution functions. In our
calculations we use CTEQ6L \cite{cteq} to generate the parton
distributions with the renormalization scale $\mu_R $ and the
factorization scale $\mu_F$ chosen to be $\mu_R = \mu_F = 2m_{h}$
and the two-loop running coupling constant $\alpha_{s}$ with
$\alpha_{s}(m_{Z})=0.118$.  The SM parameters involved are taken as
 $m_t=172.7$ GeV \cite{top-mass} and $m_{Z}=91.187$ GeV \cite{ref16}.
We fix  $m_{h}=150$ GeV in our numerical calculations.
The new free parameters involved are the
symmetry breaking scale $f$, the ratio $r=\lambda_1/\lambda_2$,
$\kappa$, $m_q$ and $m_\chi$. Our calculations show that the results
are not sensitive to $\kappa$, $m_q$ and $m_\chi$ for $m_q,m_\chi  >
3$ TeV, which is in agreement with the finding in \cite{ref11}.
Thus, we take $\kappa=1.0$, $m_q=m_\chi=5$ TeV and retain $f$ and $r$
as free parameters.

In Fig. 2 we plot the hadronic Higgs-pair production cross section
at the LHC versus the parameter $f$ for several values of $r$.
Here, we included all effects from the top-quark,  T-even and
T-odd quarks (three generations).
Fig. 2 shows that the contributions of this model
increase the SM cross section in the allowed parameter space, and
the magnitude of such corrections depends on the parameters $r$ and
$f$. The corrections are sensitive to the scale $f$ and become more
sizable for lower values of $f$. For example, for $r=0.5$, the
total cross section can reach $30$ fb.
In Fig. 2 we also show a typical prediction by supersymmetric model
from  the first reference of \cite{ref6}.
Note that in the minimal supersymmetric model the Higgs boson mass $m_h$
is upper bounded by 135 GeV and cannot be as heavy as 150 GeV which we
choose for both the SM and the littlest Higgs model. We see that
the production rate in the littlest Higgs model with T-parity
can be larger than the supersymmetric model prediction in the allowed
parameter space.

\begin{figure}[tb]
\begin{center}
 \epsfig{file=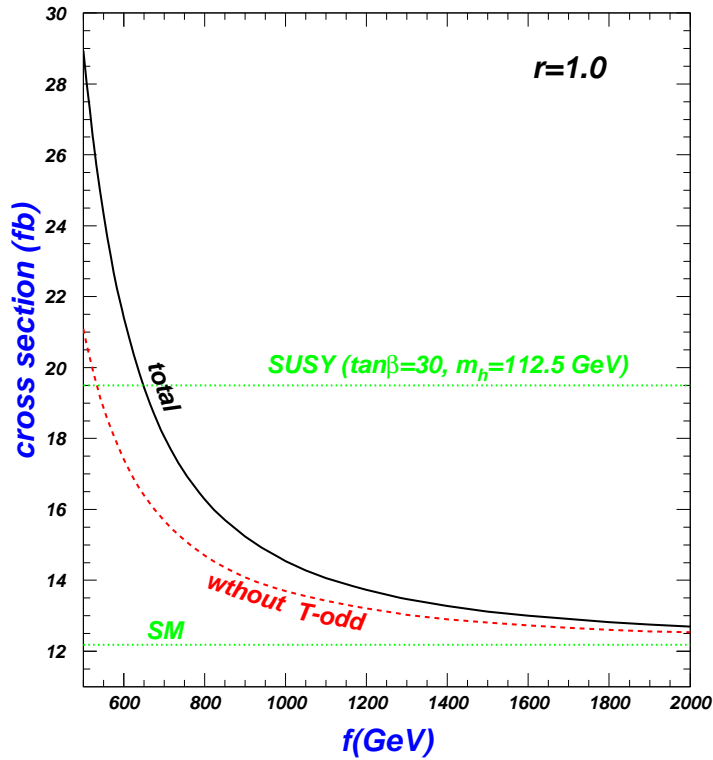,width=10cm}
\end{center}
\vspace{-1.0cm}
\caption{Same as Fig.2, but show the results with and without the
contributions of the T-odd quarks for $r=1.0$. }
\end{figure}

The comparison of the results with and without T-odd
quark contributions is shown in Fig. 3. We see that
the contributions of T-odd quarks are equally important and
thus cannot be neglected.
For example, with (without) the contributions of T-odd quarks,
the cross section is 29 fb (21 fb).

\section{Conclusion}
In the framework of the littlest Higgs model with T-parity
we calculate the production of a pair of neutral CP-even Higgs bosons
at the LHC.
We found that, due to the loop contributions of both T-even and T-odd quarks
predicted in this model, the production rate can be significantly enhanced
relative to the Standard Model prediction.
Also, we found that the T-odd quark contributions, which were ignored in
previous studies, are equally important compared with the T-even quark
contributions and thus cannot be negelected.

\end{document}